\documentclass{aa}

\def\puncspace{\ifmmode\,\else{\ifcat.\C{\if.\C\else\if,\C\else\if?\C\else%
\if:\C\else\if;\C\else\if-\C\else\if)\C\else\if/\C\else\if]\C\else\if'\C%
\else\space\fi\fi\fi\fi\fi\fi\fi\fi\fi\fi}%
\else\if\empty\C\else\if\space\C\else\space\fi\fi\fi}\fi}
\def\SP{\let\\=\empty\futurelet\C\puncspace}

\def\h1{$h^{-1}$\SP}

\def\etal{{\it et al.\/}\ }

\def\eg{{\it e.g.\/}\rm,\ }
\def\lsim{~\rlap{$<$}{\lower 1.0ex\hbox{$\sim$}}}
\def\gsim{~\rlap{$>$}{\lower 1.0ex\hbox{$\sim$}}}
\def\void#1{{}}

\usepackage{graphics}
\begin{document}
 \thesaurus{(11.03.1); (12.12.1); (12.03.3)}

\title{VLT and NTT Observations of Two EIS Cluster Candidates. }
\subtitle{Detection of the Early-Type Galaxies Sequence at
$z\sim1$\thanks{Based on observations collected at the European
Southern Observatories of Paranal (VLT-UT1 Science Verification
Program) and La Silla (Chile).}}

\author{} \author{L. da Costa\inst{1} \and M. Scodeggio\inst{1} \and
L.F. Olsen\inst{1,2} \and M. Nonino\inst{1,3} \and
R. Rengelink\inst{1} \and R. Bender\inst{4} \and M. Franx\inst{5} \and
H.E. J{\o}rgensen\inst{2} \and A. Renzini\inst{1} \and P. Rosati\inst{1} }
\offprints{L. da Costa} \institute{ European Southern Observatory,
Karl-Schwarzschild-Str. 2, D--85748 Garching b. M\"unchen, Germany
 \and Astronomisk Observatorium, Juliane Maries Vej 30, DK-2100
Copenhagen, Denmark \and Osservatorio Astronomico di Trieste, Via G.B.
Tiepolo 11, I-31144 Trieste, Italy  \and
Universit\"ats-Sternwarte M\"unchen, Scheinerstr. 1, D-81679,
M\"unchen, Germany \and Leiden Observatory, P.O. Box
9513, NL-2300 RA, Leiden, The Netherlands  }

\date{Received ; accepted}
\maketitle

\begin{abstract}

Optical data from the ESO VLT-UT1 Science Verification observations
are combined with near-infrared data from SOFI at the NTT to obtain
optical-infrared color-magnitude diagrams for the objects in the
fields of two EIS cluster candidates. In both cases, evidence is found
for a well-defined sequence of red galaxies that appear to be
significantly more clustered than the background population. These
results suggest that the two systems are real physical
associations. The $(R-Ks)$, $(I-Ks)$ and $(J-Ks)$ colors of the red
sequences are used, in conjunction with similar data for
spectroscopically confirmed clusters, to obtain redshift estimates of
$z\sim 0.9$ and $z\sim 1.0$ for these two systems. These results make
these EIS cluster candidates prime targets for follow-up spectroscopic
observations to confirm their reality and to measure more accurately
their redshift.

\end{abstract}

\section{Introduction}
\label{intro}

Growing evidence for the existence of clusters at $z\sim1$ and beyond
makes the identification and study of these systems of great interest
for probing galaxy evolution and cosmological models.  However, the
number of confirmed systems at these high redshifts is currently very
small, precluding any robust statistical analysis. The largest sample
of spectroscopically confirmed clusters has been selected from ROSAT
deep X-ray observations (Rosati \etal 1998, Rosati 1998), while a few
other $z\sim 1$ clusters have been discovered in the surroundings of
strong radio sources (\eg Dickinson 1996; Deltorn \etal 1997), or
using infrared observations (\eg Stanford \etal 1997). Although X-ray
and infrared searches are very effective in identifying real clusters,
their ability to cover large areas of the sky is presently limited,
and these methods are not likely to produce large samples of very
distant clusters in the short-term. On the other hand, with the advent
of panoramic CCD imagers, optical wide-angle surveys have become
competitive in identifying cluster candidates up to $z\sim1$. Examples
of such surveys, suitable for cluster searches, include those of
Postman \etal (1996), Postman \etal (1998) and the ESO Imaging Survey
(EIS, Renzini \& da Costa 1997), which cover 5, 16 and 17 square
degrees, respectively, reaching $I_{AB} \lsim24$. These surveys are
currently being used for systematic searches of galaxy cluster
candidates employing objective matched-filter algorithms (\eg Postman
\etal 1996).  In the case of the EIS project, about 300 candidates
have been identified, over the redshift interval $0.2 \lsim z \lsim
1.3$, out of which 79 are estimated to have $z\gsim0.8$ (Olsen \etal
1998a, b; Scodeggio \etal 1998). However, only with additional
observations can these optically-selected high-redshift candidates be
confirmed. Establishing the global success rate of this technique (and
its possible redshift dependence) is extremely important for the
design of future wide-field optical imaging surveys. Indeed, these
surveys may play a major role in significantly increasing the number
of known distant clusters, thus making them useful tools for probing
the high-$z$ universe.

As a test case, two EIS cluster candidates identified in EIS patch B
(EIS~0046-2930 and EIS~0046-2951; Olsen \etal 1998b), were observed
with the VLT Test Camera (VLT-TC) as part of the ESO VLT-UT1 Science
Verification (SV; see Leibundgut, De Marchi \& Renzini 1998).
After the public release of these Science Verification data, fields
including the two candidate clusters have been observed at the ESO
3.5m New Technology Telescope (NTT), as part of an ongoing infrared (IR)
survey of EIS patch~B (J{\o}rgensen \etal 1999). Therefore, we had the
opportunity to combine the VLT optical data with the NTT IR
data, and to use both optical and IR color-magnitude (CM)
diagrams to search for evidence of a ``red sequence''
of luminous early-type galaxies, typical of populous clusters at low
as well as at high redshift (\eg Bower, Lucey \& Ellis 1992; Stanford,
Eisenhardt \& Dickinson 1998; Kodama \etal 1998).  A clear
identification of this sequence would provide strong support to the
reality of the clusters, while allowing an independent estimate of
their redshift to be obtained.

In this Letter we briefly describe the various observations and the
data reduction in Section 2; in Section 3 we present our results; and
in Section 4 we summarize our conclusions.

\section {Observations and Data Reduction}
\label{sec:obs}

Originally, four EIS cluster candidates were selected for the VLT-UT1
SV program, after visual inspection of all candidates found in the
EIS-wide Patch B.  The four targets were selected to cover a range in
redshift and richness among the EIS candidates. However, due to time
and weather constraints only two fields were actually observed.

\begin{table}
\caption{Summary of VLT-TC and SOFI Observations}
\label{tab:obs}
\begin{tabular}{lccc} 
\hline\hline
Cluster   &   Filter &   $t_{int}$   &   seeing \\
          &          &    (seconds)          & arcsec \\
\hline
EIS 0046-2930 &  $V$  &  2700           &   0.9       \\
              &  $R$  &  2700           &   0.8       \\ 
              &  $I$  &  1500             &   0.9       \\
              &  $J$  &  3000           &   0.8       \\
              &  $Ks$ &  2400          &   1.1       \\
EIS 0046-2951 &  $V$  &  2700            &   1.1       \\
              &  $I$  &  1600              &   0.8       \\
              &  $J$  &  3000              &   1.3       \\
              &  $Ks$ &  3000             &   0.9       \\
\hline\hline
\end{tabular}
\end{table}

The optical observations were conducted on the nights of August 18 and
23, 1998 with the Test Camera of the VLT-UT1, as part of the ESO
VLT-UT1 Science Verification (1998). The VLT-TC was equipped with an
engineering grade Tektronix $2048^2$ CCD, covering a field of view of
about 93 $\times$ 93 arcsec with an effective pixel size of 0.09
arcsec (after a $2 \times 2$ rebinning).  One of the cluster
candidates (EIS~0046-2930) was observed in $VRI$, and the other
(EIS~0046-2951) only in the $V$ and $I$ passbands. In
Table~\ref{tab:obs} we summarize the available data, giving the
passband, the corresponding total integration time and the median
seeing of the combined images. During the exposure of EIS~0046-2930
the transparency was poor and variable, leading to fairly bright
limiting magnitudes.  Single exposures have been reduced by the ESO
Science Verification Team using standard {\sl IRAF} procedures, and
then publicly released.  These reduced images were then processed
using the EIS pipeline which performed the astrometric and photometric
calibration, and coaddition for each band (see Nonino \etal 1998). The
VLT-TC optical data were calibrated against the EIS data, for which
the uncertainty in the photometric zero-point was estimated to be 0.1
mag in V and 0.02 mag in I. The VLT-TC versus EIS comparison yields an
additional uncertainty of about 0.1 mag.  Therefore, we estimate that
the overall uncertainty in the zero-points is $\lsim 0.15$~mag in $V$
and $\lsim$ 0.12 in $I$.

\begin{table*}
\caption{Cluster Properties}
\label{tab:clusters}
\begin{tabular}{lcccccccc} 
\hline\hline
Cluster  &  $\alpha_{EIS}$ & $\delta_{EIS}$ &  $\alpha_{new}$ & 
$\delta_{new}$ &$\sigma$ & $N_R$ & $z_{EIS}$ &$z_{CM}$ \\
\hline
 EIS 0046-2930 & 00:46:29.6 & -29:30:57 & 00:46:27.6 & -29:30:38 & 3.0
&  46     &  0.6    & 1.0 $\pm$ 0.1   \\
 EIS 0046-2951 & 00:46:07.4 & -29:51:44 & 00:46:06.7 & -29:51:26 & 3.2
&   2     &  0.9    & 0.9 $\pm$0.15  \\
\hline\hline 
\end{tabular}
\end{table*}

The IR $J$ and $Ks$ band images were obtained on October 8 and
9, 1998 using the SOFI infrared spectrograph and imaging camera
(Moorwood, Cuby \& Lidman 1998) at the NTT.  SOFI is equipped with a
Rockwell 1024$^2$ detector that, when used together with the large
field objective, provides images with a pixel scale of 0.29 arcsec,
and a field of view of about $4.9 \times 4.9$ arcmin.  The full set of
SOFI observations will be described elsewhere (J{\o}rgensen \etal
1999); here we describe only those for the fields including the two
cluster candidates.  Total integration times and the seeing measured
on the combined images are given in Table~\ref{tab:obs}. The data were
reduced using the Eclipse data analysis software package (Devillard
1998), developed to combine jittered images. The resulting combined
images were then input to the EIS pipeline for astrometric and
photometric calibrations using observations of standard stars given by
Persson (1997). From the scatter of the photometric solution we
estimate the zero-point uncertainty in the $J$ and $Ks$ bands to be
$\lsim 0.1$ mag.  
\void{We point out that our conclusions are not
significantly affected by the uncertainties of the optical and
infrared zeropoints.}

In order to facilitate the analysis of the whole dataset, the images
from the EIS-wide survey were resampled to a common reference frame,
centered on the initial estimate of the two candidate cluster
positions, using the Drizzle routine of the EIS pipeline.  The
resampled images have the same pixel size as the SOFI images.  To
improve the sensitivity to faint objects the resampled EIS-wide and
SOFI images were combined to produce one very deep $B + V + I + J +
Ks$ image for each field. This image has a sufficiently large field of
view ($4.9 \times 4.9$ arcmin) to allow a reliable estimate of the
background source density to be obtained (see Section~3). The source
extraction software SExtractor (Bertin \& Arnouts 1996) was
subsequently used to detect sources in these deep images, while
measuring the flux parameters for each individual passband in the
separate images. Magnitudes and colors were measured using a 4 arcsec
diameter aperture.

\begin{figure*}
\resizebox{\textwidth}{!}{\includegraphics{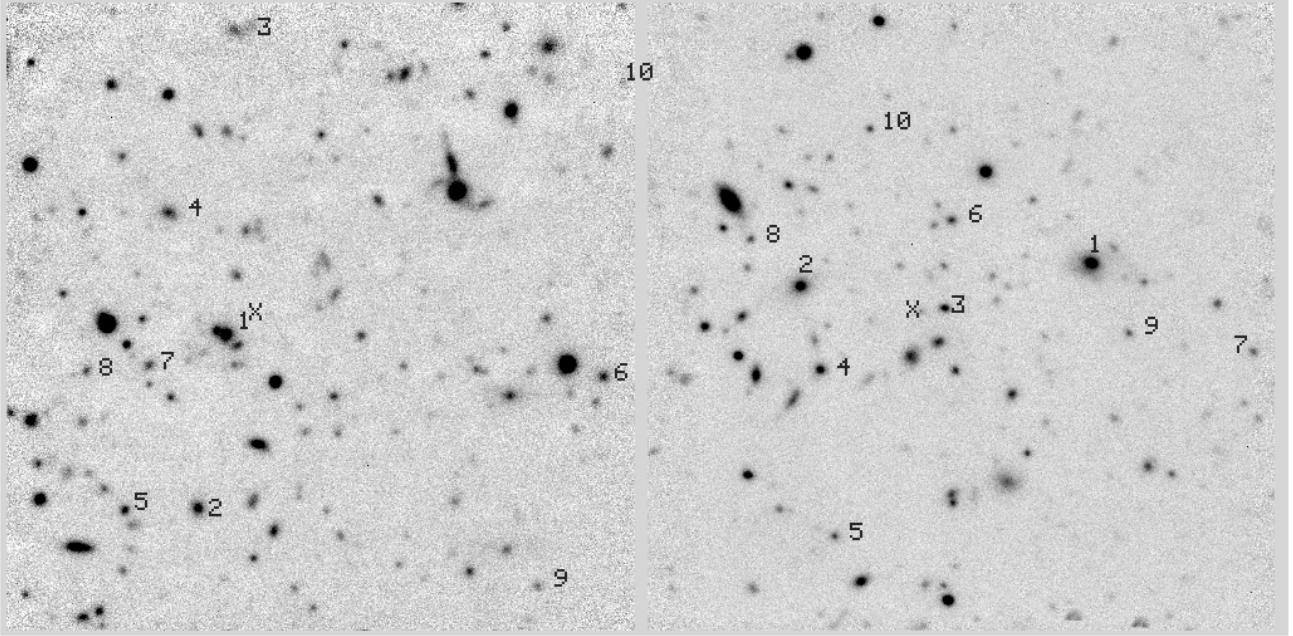}}
\caption{Combined VLT-TC images (see text) for EIS~0046-2930 ($V+R+I$,
left panel) and EIS~0046-2951 ($V+I$, right panel). The field of view
of these images is $\sim$ 90 arcsec. The computed flux-weighted
centers are marked by $\times$. The numbers indicate the magnitude
ranking of the ten brightest, most likely cluster members as described
in the text.}
\label{fig:images}
\end{figure*}

Also all available VLT images were coadded to produce the $V + R + I$
and $V + I$ images shown in Figure~\ref{fig:images}.  The resulting
images are considerably deeper than those from EIS, and also have much
better resolution.  This procedure has allowed us to reach
approximately the same limiting magnitude in both fields ($I\sim 25.0$
at about $2\sigma$). Even though the transparency during the
observations of the EIS~0046-2930 field was poor, this was
compensated by the addition of the $R-$band exposure. We have also
resampled and combined the VLT-TC images with those from SOFI, using
the same method as above.

Using the available multicolor data from EIS-wide plus SOFI, we
derived $(I-Ks)-Ks$ CM-diagrams for all galaxies within 1 arcmin of
the nominal candidate cluster centers. From these diagrams, a
tentative color-based selection was made, dividing galaxies into
cluster candidate members and foreground/background objects. Based on
this selection, we computed for each cluster candidate a new position,
obtained as the flux-weighted center of mass of the candidate
members. An identical procedure was carried out using the SOFI data
only, leading to very similar results. In both cases the new positions
were found to be within 0.4 arcmin of the position given in the EIS
catalog (Olsen \etal 1998b). Note that this corresponds to the pixel
size (0.45 arcmin) of the maximum-likelihood map used in the EIS
cluster finding procedure.

\section{Results}

Table~\ref{tab:clusters} gives the original cluster candidates
coordinates, the new flux-weighted positions as described above, the
significance of the detection, the Abell richness and the estimated
redshift from the EIS catalog, as listed in Olsen \etal (1998b) and
the new redshifts derived below using the CM-diagrams from the
combined VLT-TC and SOFI data. All coordinates are in J2000.  Note
that the estimate of the Abell richness for distant clusters is quite
uncertain, but it serves to indicate their relative richness.

\begin{figure}
\resizebox{\columnwidth}{!}{\includegraphics{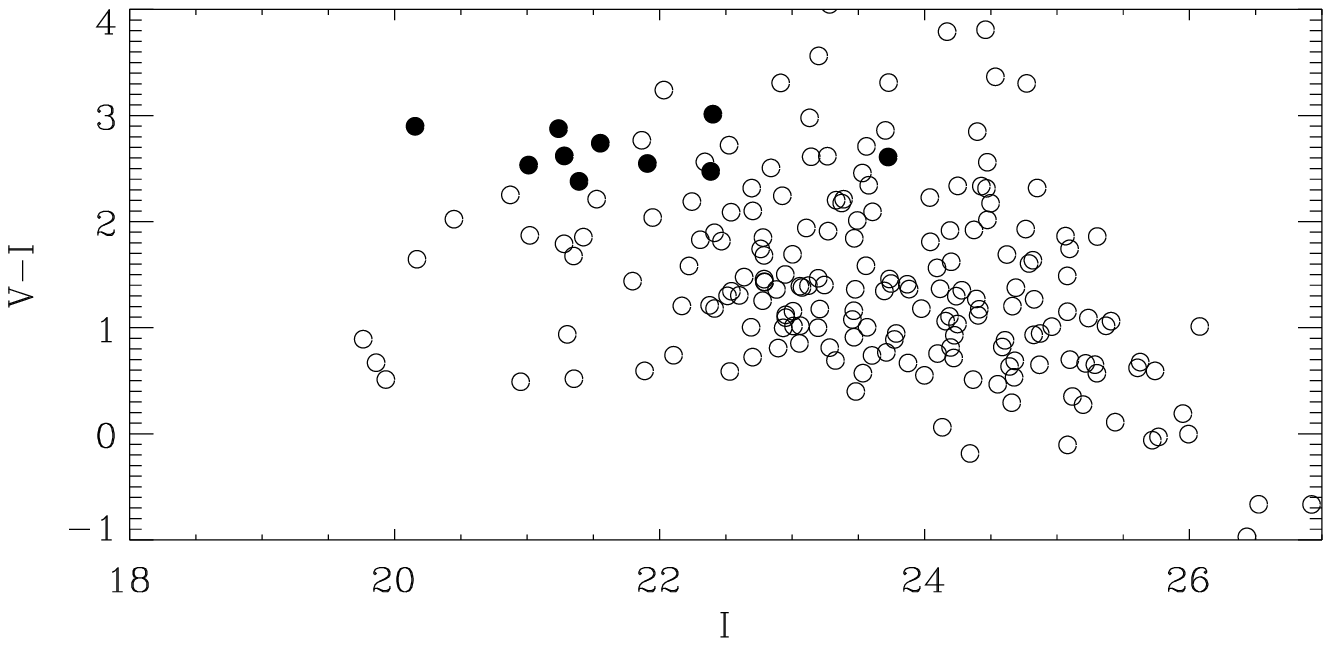}}
\resizebox{\columnwidth}{!}{\includegraphics{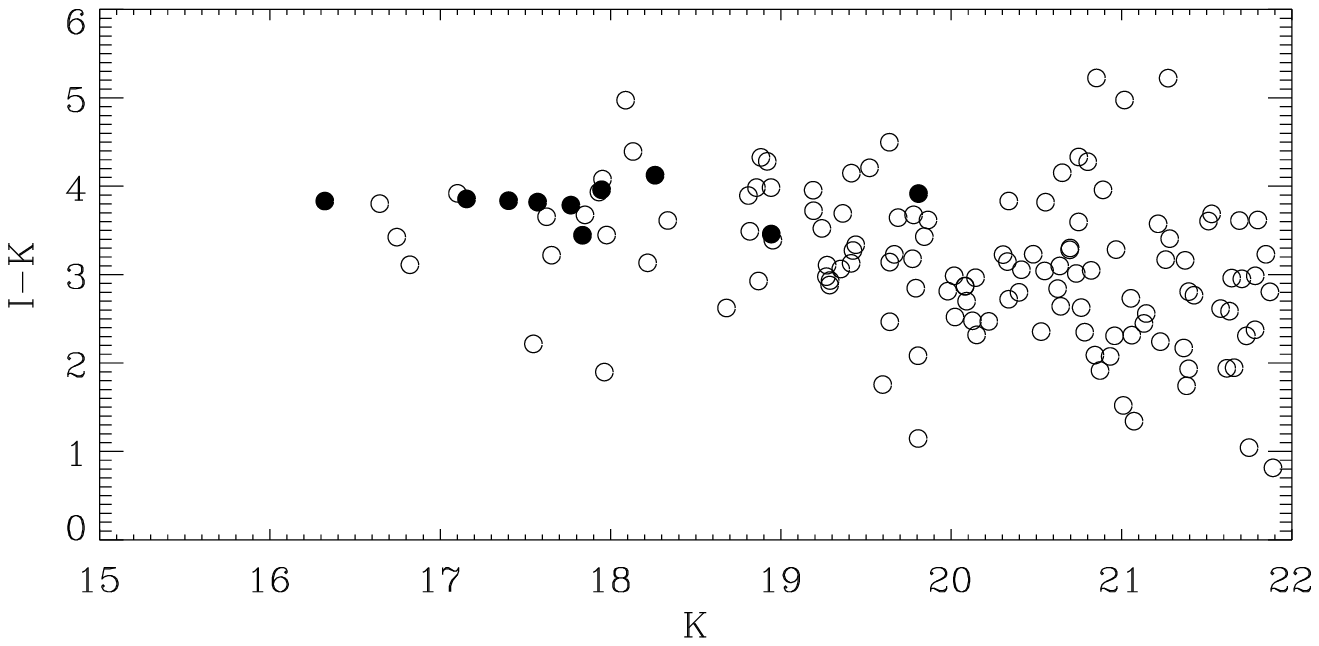}}
\resizebox{\columnwidth}{!}{\includegraphics{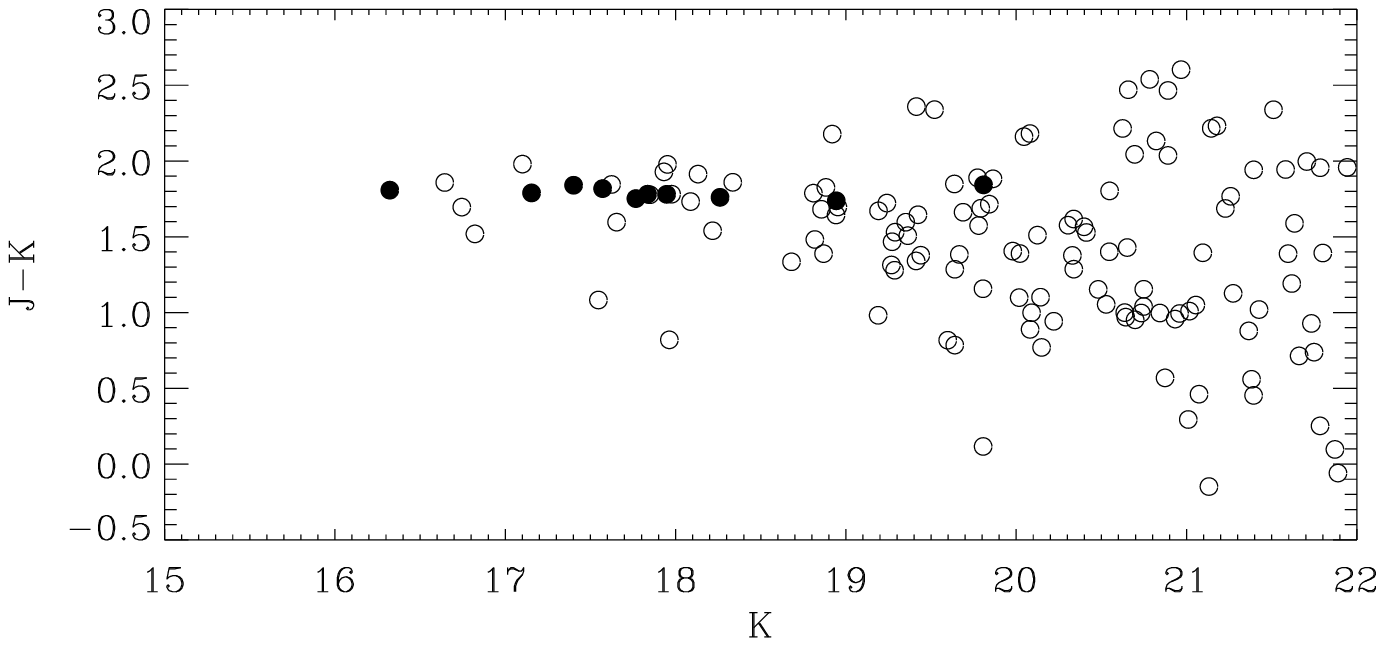}}
\resizebox{\columnwidth}{!}{\includegraphics{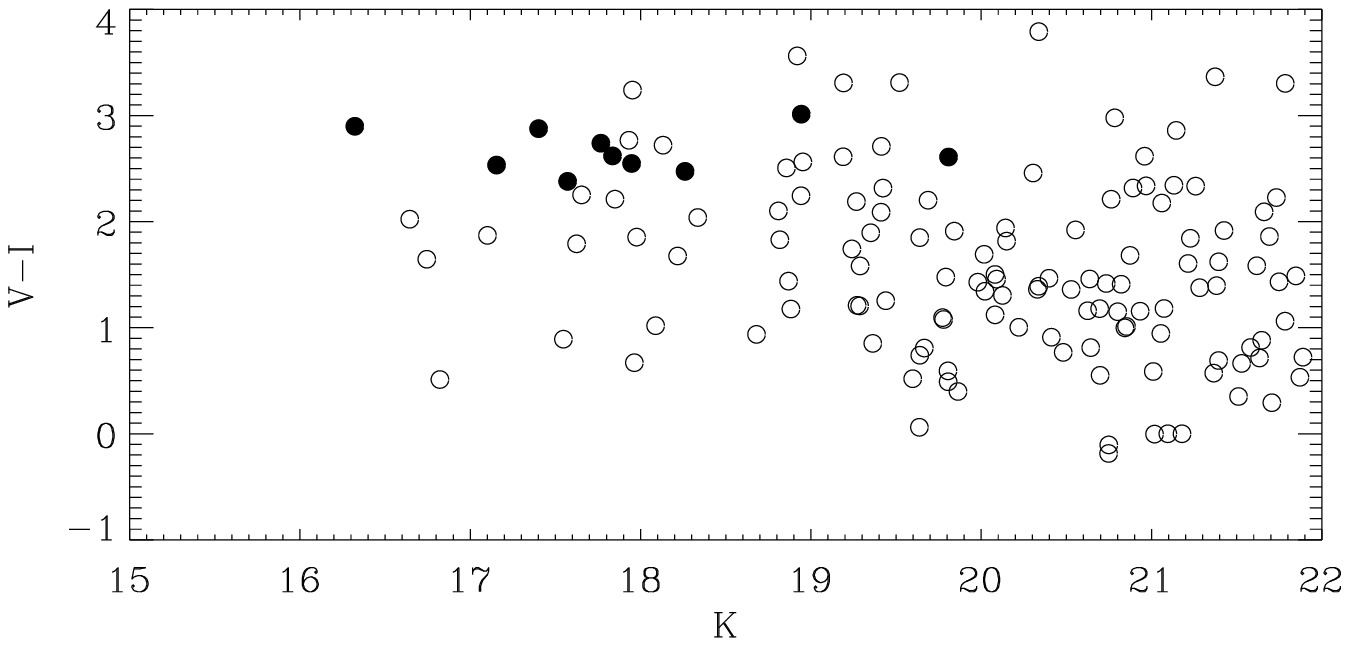}}
\caption{Optical and infrared CM-diagrams for all galaxies in the
VLT-TC field of EIS~0046-2930. The filled circles indicate the ten
brightest, most likely cluster members based on a color selection as
described in the text. These galaxies are also marked in
Figure~\protect\ref{fig:images}.}
\label{fig:cmb18cm}
\end{figure}
\begin{figure}
\resizebox{\columnwidth}{!}{\includegraphics{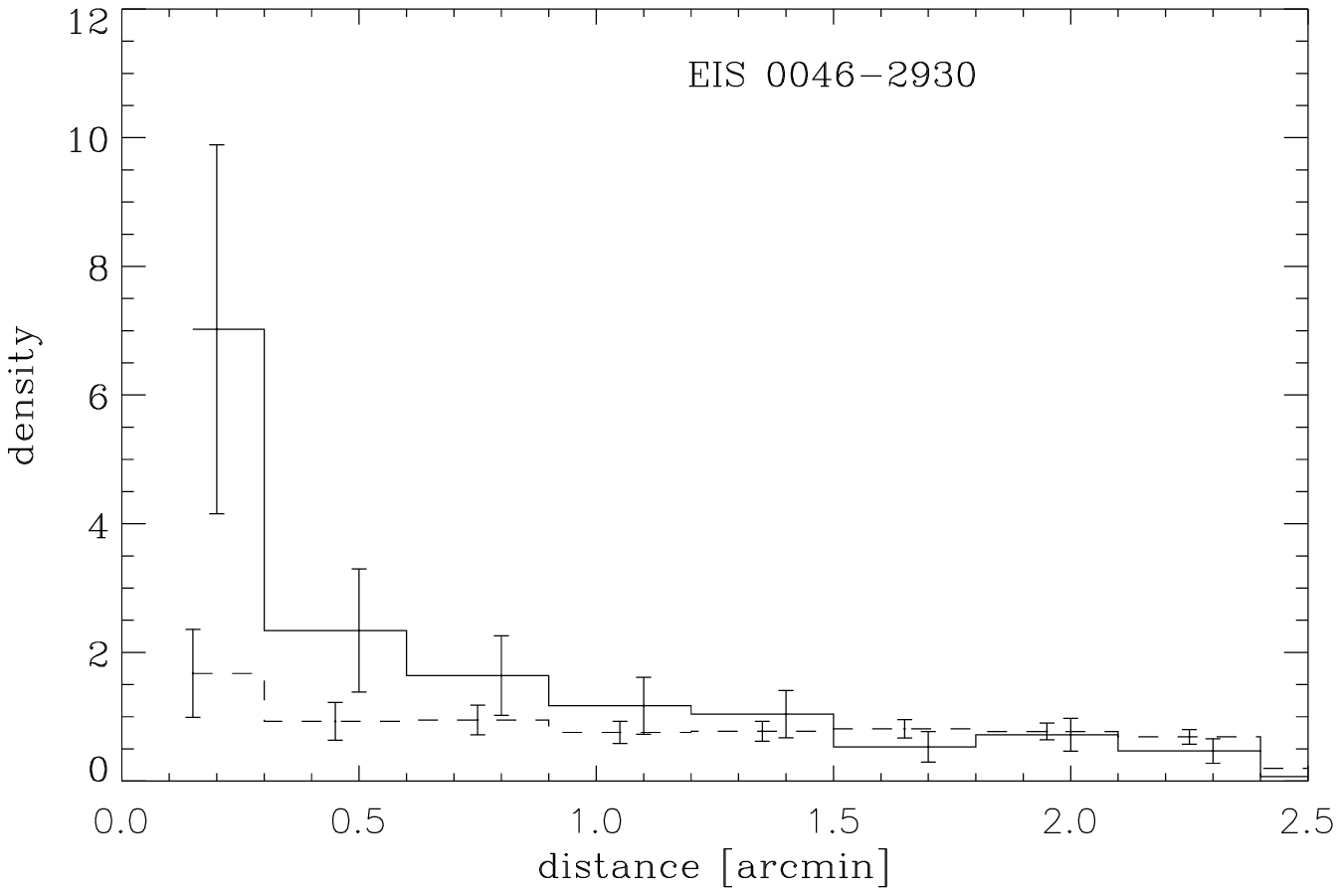}}
\caption{Projected radial distribution of galaxies brighter than
$Ks=20$ within the SOFI field of EIS~0046-2930. The figure shows
separately the distribution of galaxies within $1.7<(J-Ks)<1.9$ (solid
line) and outside this color range (dashed line), both normalized to
their respective backgrounds.}
\label{fig:profb18}
\end{figure}

\begin{figure}
\resizebox{\columnwidth}{!}{\includegraphics{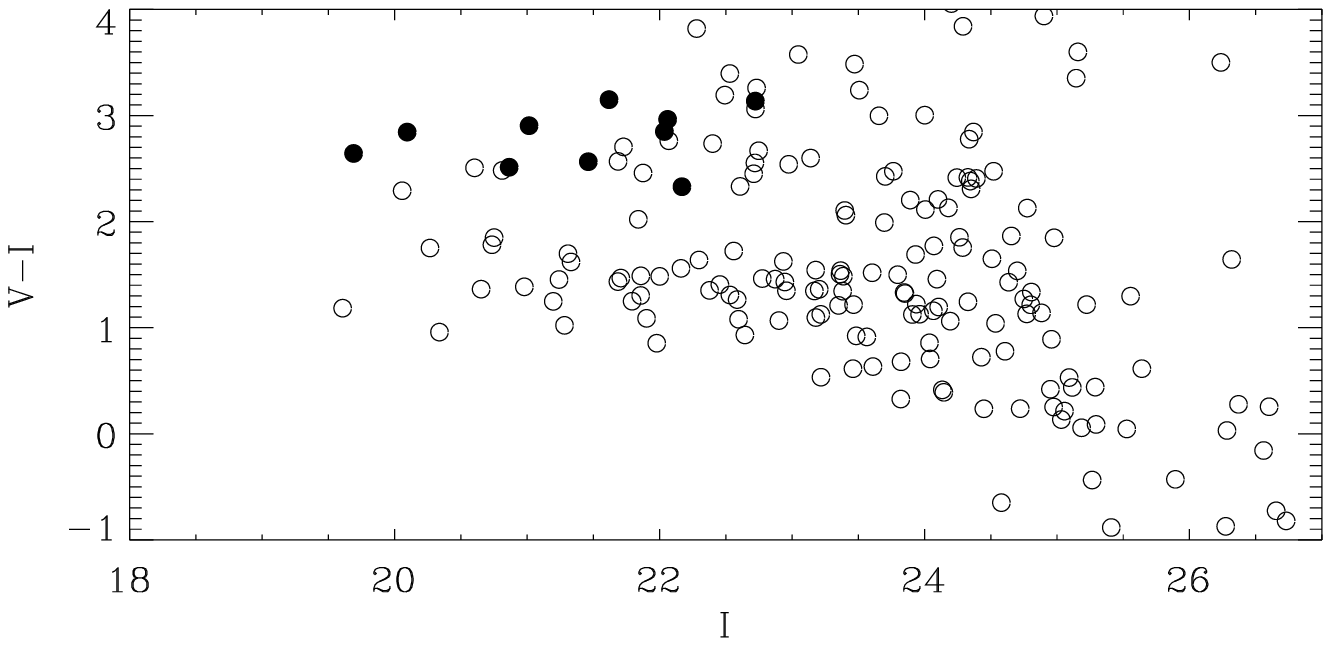}}
\resizebox{\columnwidth}{!}{\includegraphics{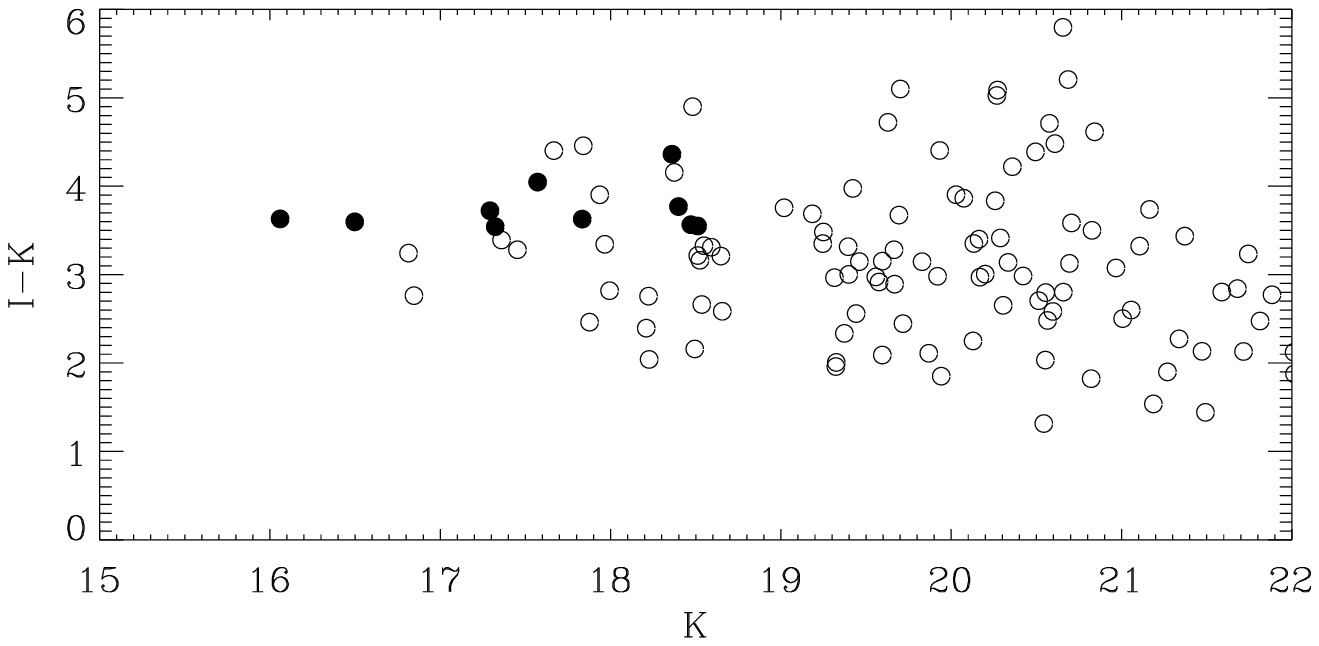}}
\resizebox{\columnwidth}{!}{\includegraphics{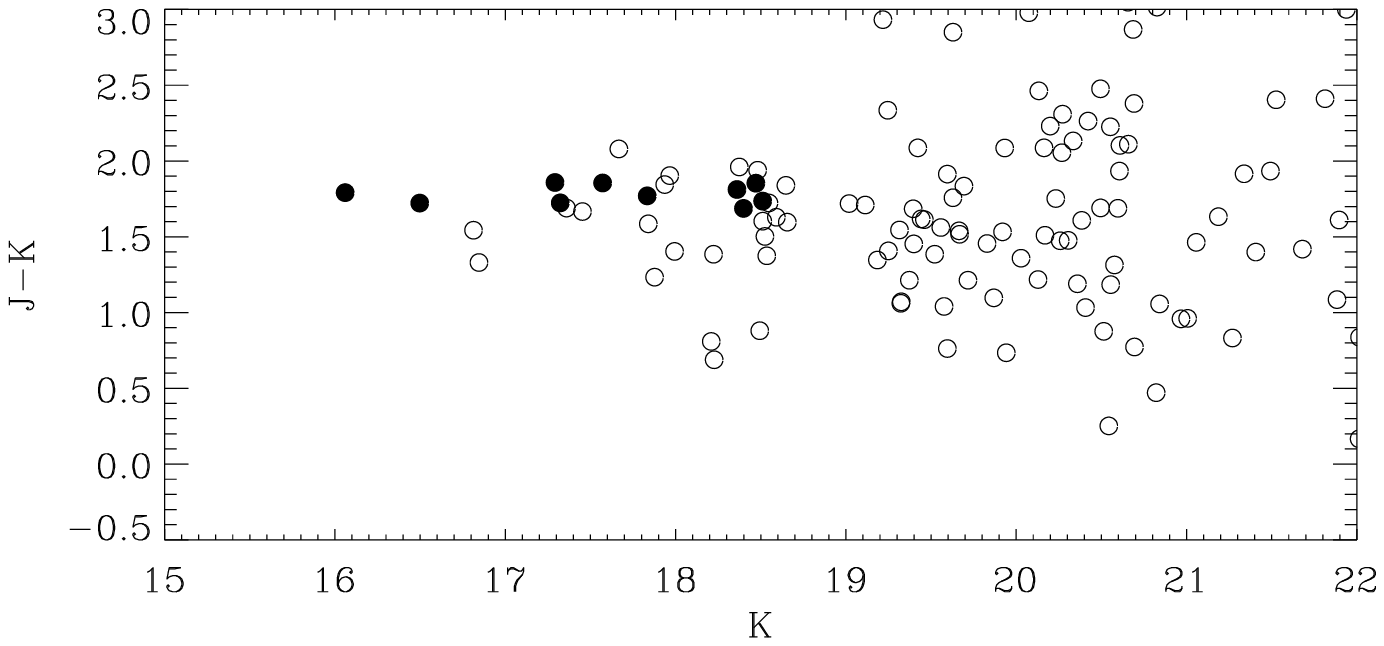}}
\resizebox{\columnwidth}{!}{\includegraphics{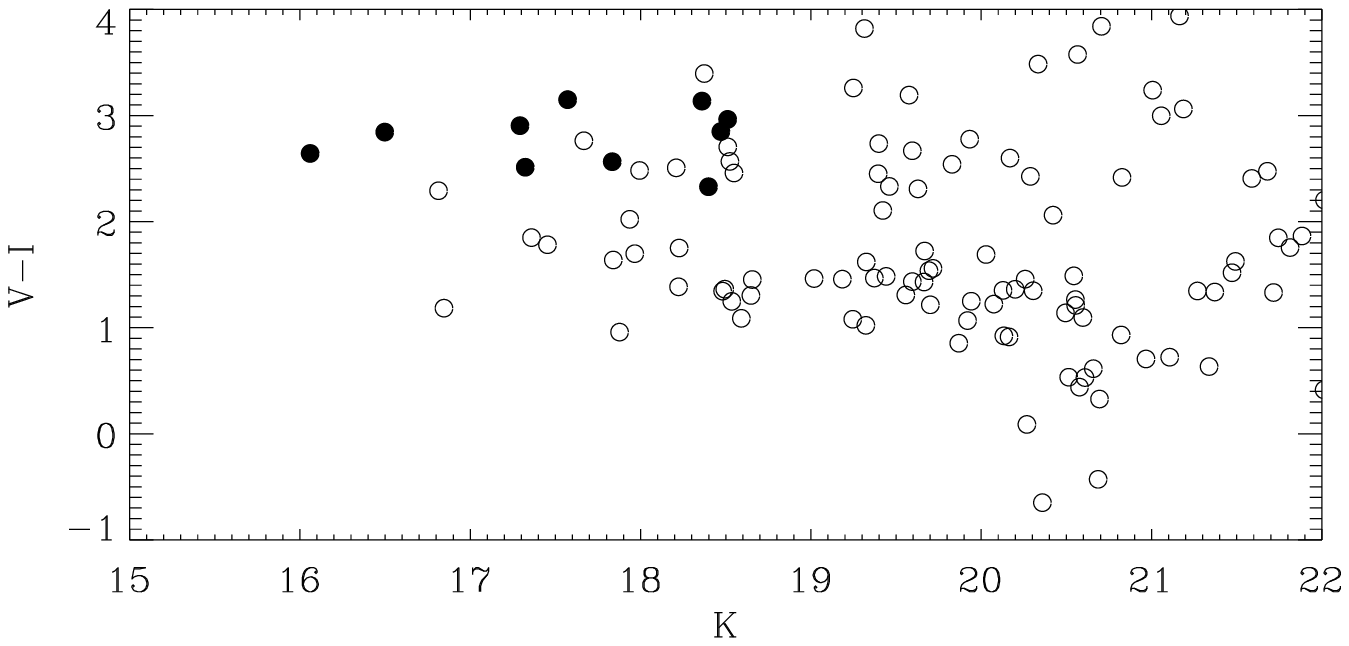}}
\caption{Optical and infrared CM-diagrams for galaxies in the VLT-TC
field of EIS 0046-2951. The filled circles represent the ten
brightest, most likely cluster members (marked in
Figure~\protect\ref{fig:images}) as described in the text. }
\label{fig:cmb5cm}
\end{figure}

\begin{figure}
\resizebox{\columnwidth}{!}{\includegraphics{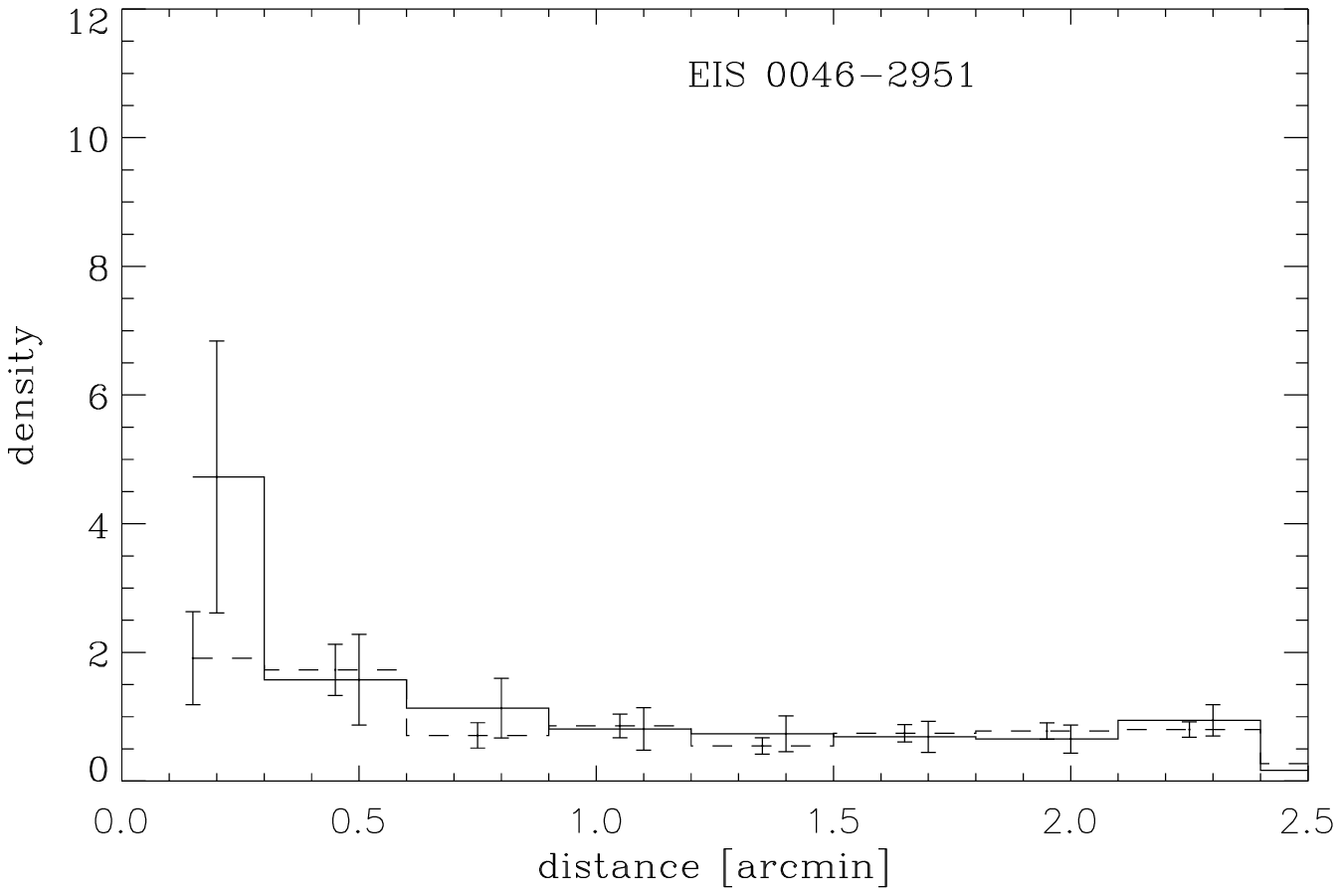}}
\caption{Projected radial distribution of galaxies brighter than
$Ks=20$ within the SOFI field of EIS~0046-2951. The figure shows
separately the distribution of galaxies within $1.6<(J-Ks)<1.8$ (solid
line) and outside this color range (dashed line), both normalized to
their respective backgrounds.}
\label{fig:profb5}
\end{figure}

\subsection{EIS 0046-2930}

In the EIS candidate clusters catalog this object was identified only
in $I$-band, and assigned a redshift of $z_{EIS} \sim 0.6$. However,
visual inspection of the original survey images of this field showed
the presence of foreground ``blue'' galaxies and of a fainter red
population, not detected in the $V-$band.  Using the deeper optical
(reaching $V\sim26.0-26.5$) and the IR catalogs produced from
the VLT-TC and from the SOFI images, we can study in greater detail
this cluster candidate. The resulting four optical
and IR CM-diagrams are shown in Figure~\ref{fig:cmb18cm}, for
all galaxies within the VLT-TC field of view.  The upper panel shows
the optical $(V-I)-I$ CM-diagram, where there is a suggestion for a
concentration of galaxies at $(V-I) \sim 2.7$, just beyond the reach
of the EIS color data.  However, the scatter is large compared with
that seen in clusters at intermediate redshifts (Olsen \etal 1998b),
and cannot be explained by photometric errors in the color which at
$I\sim 22$ are $\lsim$ 0.3 mag. This scatter prevents a secure
identification of the red sequence. By contrast, the $(I-Ks)-Ks$
diagram shows a clear early-type sequence in the interval
$Ks=16.0-20.0$ at $(I-Ks) \sim 3.9$. Using the above magnitude range,
the CM-relation is well-fitted by a linear relation with an estimated
scatter of $\sim 0.15$ ($Ks<18.5$), comparable to the estimated error
in the color and in agreement with the color dispersion of
morphologically classified early-type galaxies in high-$z$ clusters
(Stanford, Eisenhardt \& Dickinson 1998).  The infrared $(J-Ks)-Ks$
diagram shows an even tighter sequence at $(J-Ks)\sim1.8$, with a
scatter of $\sim$0.1 mag, again comparable with the estimated error in
the color. The ten brightest galaxies (in the $Ks$ band) for which
$1.7\le (J-Ks) \le 1.9$ {\it and} $(V-I)\geq 2.3$ are represented by
filled circles in the CM-diagrams and are also numbered in
Figure~\ref{fig:images} according to their magnitude ranking.  These
objects are the most likely early-type galaxy members of
EIS~0046-2930. The flux-weighted position of the ``cluster'' is also
shown.

The projected radial distribution of objects brighter than $Ks=20$ and
within the color range $1.7\lsim(J-Ks)\lsim1.9$, is shown in
Figure~\ref{fig:profb18}, in annuli 0.3 arcmin wide. The contrast of
this bright red-sequence population relative to the background is
clearly seen, while there appears to be no appreciable clustering for
galaxies outside this color range. Even though the statistic is poor,
the scale and amplitude of the overdensity associated to this
population, a factor of $\sim$ 7 within the innermost 0.3 arcmin, are
similar to those observed by Dickinson (1996) for the cluster
surrounding 3C~324 at $z\simeq 1.26$.  To test the robustness of this
results flanking fields with the same size of the VLT-TC field of view
were extracted from the same SOFI image and used to obtain CM-diagrams
and radial density profiles. None of these fields showed the presence
of a concentration of galaxies both in color and in position.  This
suggests that the concentration of galaxies in both color and
projected separation seen in the field of EIS~0046-2930 is significant
and that this object is likely to be a real cluster.
Further support to this conclusion comes from the matched-filter
algorithm which applied to the $Ks-$band data detects a ``cluster'' at
the $3\sigma$ level and at $z\sim 1$.  

On the presumption that EIS~0046-2930 is a real cluster, the color of
the red sequence can be used to estimate its redshift.  This can be
achieved either by using synthetic stellar population models, or
purely empirically using the colors of the red sequence of clusters of
known redshift. Even though the available data are sparse, we have
adopted the latter approach because it is model independent. We have
used the spectroscopic redshifts and the CM-diagrams given by
Stanford, Eisenhardt \& Dickinson (1998) for their clusters at $z
\gsim 0.5$ and the $z=1.273$ cluster of Stanford \etal (1997) to
estimate the location of the early-type galaxies sequence in different
passbands for clusters at $z\sim1$. Interpolating these relations to
the colors of the red sequence of EIS~0046-2930 ($(R-K)=5.4$,
$(I-K)=3.9$, and $(J-K)=1.8$) we consistently estimate its redshift to
be $z_{CM} = 1.0\pm 0.1$ (statistical uncertainty only).

\subsection {EIS 0046-2951}

In the EIS catalog this object was estimated to have a redshift of
$z_{EIS} \sim 0.9$, being detected only in the $I$-band
(Table~\ref{tab:clusters}). However, visual inspection of the $V-$ and
$I-$band EIS images suggested that this system could be an overlap of
two concentrations at different redshifts. Using the deeper $V-$band
image obtained with the VLT-TC we are now able to investigate the
optical CM-diagram shown in Figure~\ref{fig:cmb5cm}. Indeed, we find
two concentrations of galaxies: one seen at $(V-I) \sim 1.6$ and
another at $(V-I) \sim 2.6$. These colors correspond to redshifts
$z\sim0.25$ and $z\sim0.7$, respectively. However, in the $(I-Ks)-Ks$
and $(J-Ks)-Ks$ CM-diagrams only one sequence is seen, located at
$(I-Ks)\sim3.5$ and $(J-Ks)\sim1.7$. These values lead to redshift
estimates of $z_{CM}=0.9 \pm 0.15$ in both cases, in good agreement
with the original estimate based on the matched-filter algorithm.  In
contrast to the previous cluster, the scatter of the red sequence in
both colors is significantly larger (0.21 in $(I-Ks)$ and 0.19 in
$(J-Ks)$) and cannot be fully accounted for by the photometric errors
in our data ($\lsim$ 0.15 mag). The larger scatter may be due to a
larger fraction of spiral galaxies in the ``cluster'', or to a stronger
contamination by foreground galaxies. As in the previous case, the
most likely early-type cluster galaxies have been selected adopting a
color-selection criterion similar to that described above. These
galaxies, chosen to have $1.5\leq (J-Ks)\leq 1.9$ and $(V-I)\geq 2.3$,
are identified in Figure~\ref{fig:cmb5cm} and in the right panel of
Figure~\ref{fig:images}.

Figure~\ref{fig:profb5} shows the projected radial distribution of
color-selected candidate cluster members. In this case we find that
the overdensity of the red sequence galaxies is $\sim$5, over the same
radial distance as for the previous cluster. The smaller overdensity
of this candidate cluster (and perhaps the larger fraction of spirals)
is consistent with the lower original estimate of its richness
(Table~\ref{tab:clusters}).  Note that a 3$\sigma$ detection at
approximately the same redshift was obtained applying the
matched-filter algorithm to the $Ks$ data. As for the previous object,
the analysis of flanking fields from the SOFI image gives further
support to the reality of the observed concentration in color and projected
separation, suggesting the existence of a physical association.

\section{Summary}

We have used deep $V$- and $I$-band images of two EIS cluster
candidates taken during the ESO VLT-UT1 Science Verification
observations to investigate the reality of these clusters. The VLT
data were complemented by infrared data taken with SOFI at the NTT.
Optical, IR, and optical-IR CM-diagrams have been
constructed to search for the presence of the red-sequence typical of
bright early-type galaxies in clusters.

In the case of EIS~0046-2930 we find a well-defined sequence at
$(I-Ks)\sim3.9$ and $(J-Ks)\sim1.8$. These galaxies are also
concentrated relative to the background suggesting the existence of a
cluster at $z=1.0\pm0.1$. The evidence for the other candidate,
EIS~0046-2951, is less compelling even though we find a sequence at
$(I-Ks)\sim3.5$ and $(J-Ks)\sim1.7$, leading to an estimated redshift
of $z=0.9\pm0.15$, consistent with its original estimate.  However,
the scatter in the CM-diagrams is large and the density contrast of
the ``cluster'' relative to the background smaller. In any case, a
final conclusion on whether these systems are real physical
associations at high-redshift must await spectroscopic observations.

These results demonstrate once again the importance of infrared data
in locating high-redshift clusters. However, the small field of view
of present IR detectors makes large solid angle IR surveys very
expensive in terms of telescope time. On the other hand, wide angle
optical surveys can efficiently produce a great number of high
redshift candidates, but with a major fraction of them which may turn
to be spurious after time-consuming spectroscopic follow up. The
present experiment is an attempt at exploring a hybrid approach, in
which the optically selected candidates are first imaged in the IR
before being considered for spectroscopic follow up at a large
telescope such as the VLT. Besides providing a first verification of
the candidate clusters, the IR images can then be used to search for
clusters at higher redshift. The overall efficiency of this strategy
remains to be empirically determined, e.g. for the actual complement
of ESO telescopes and instruments, and the present paper represents a
first step in this direction.

\void{very expensive in terms of telescope time, thus making blind
searches for galaxy clusters in the IR unpractical. A low-risk
alternative, at least for clusters with $z \lsim 1$, would be to use,
as done in the present paper, the position of cluster candidates found
from optical wide-angle surveys as targets for pointed IR
observations. The IR data could then be used not only to search for
the early-type sequence associated with the candidate but also to
search for clusters at higher redshifts. The strategy would be
especially appealing if our candidates were confirmed by spectroscopic
follow-up. In this case, the EIS candidate list would provide a good
starting point, even though it is too early to estimate the effective
success rate of positive detections of high-redshift ($z\lsim1$),
optically-selected, cluster candidates on the basis of the present
results. However, the effort seems worthwhile since a large sample of
clusters at $z\sim1$ is an essential step for further detailed work
using 8m class telescopes and HST.}

\begin{acknowledgements}

We acknowledge that this work has been made possible thanks to the
Science Verification Team and EIS Team efforts in making the data
publicly available in a timely fashion.  Special thanks to
R. Gilmozzi, B. Leibundgut and J. Spyromilio. Part of the data
presented here were taken at the New Technology Telescope at the La
Silla Observatory under the program ID 62.O-0514. We thank Hans Ulrik
N{\o}rgaard-Nielsen, Leif Hansen and Per Rex Christensen for allowing
us to use the data prior to publication.
 
\end{acknowledgements}


\begin{thebibliography}{}
\bibitem{bertin} Bertin, A. \& Arnouts, S., 1996, A\&AS, 117, 393
\bibitem{bower} Bower, R., Lucey, J. \& Ellis, R., 1992, MNRAS, 254, 601
\bibitem{butcher} Butcher, H. \& Oemler, A. , 1984, ApJ, 285, 426
\bibitem{deltorn} Deltorn, J.M., Le F\'evre, O., Crampton, D., Dickinson, M.,
1997, ApJ, 483, L21
\bibitem{devillard} Devillard, N., 1998, Eclipse Data Analysis Software Package
(ESO: Garching)
\bibitem{dickinson} Dickinson, M., 1996, in ``Science with the VLT'',
ed. J. Bergeron, (Kluwer:Berlin), p. 274
\bibitem{jorgensen} J{\o}rgensen \etal,  1999, {\it in preparation}
\bibitem{kodama} Kodama, T, Arimoto, N., Barger, A. J. \&
Aragon-Salamanca, A., 1998, astro-ph/9802245
\bibitem{leibundgut} Leibundgut, B., De Marchi, G., \& Renzini,
A. 1998, The Messenger, 92, 5
\bibitem{moorwood} Moorwood, A., Cuby, J.G. \& Lidman, C., 1998, The
Messenger, 91, 9
\bibitem{nonino} Nonino, M., et al. 1998, A\&A, {\it in press}
(astro-ph/9803336)
\bibitem{olsena} Olsen, L.F., et al. 1998a, A\&A, {\it in press}
(astro-ph/9803338)
\bibitem{olsenb} Olsen, L.F., et al. 1998b, A\&A, {\it submitted}
(astro-ph/9807156)
\bibitem{persson} Persson, S.E., 1997, {\it private communication}
\bibitem{postmana} Postman, M., Lubin, L.M., Gunn, J.E., Oke, J.B.,
Hoessel, J.G., Schneider, D.P., Christensen, J.A. 1996, AJ, 111, 615
\bibitem{postmanb} Postman, M., Lauer, T.R., Szapudi, I., Oegerle, W.,
1998, ApJ, 506, 33
\bibitem{} Renzini, A. \& da Costa, L., 1997, The Messenger, 87, 23
\bibitem{rosatia} Rosati, P., Della Ceca, R., Burg, R., Norman, C., \&
Giacconi, R. 1998, ApJ, 445, L11
\bibitem{rosatib} Rosati, P., 1998, astro-ph/9810054
\bibitem{scodeggio} Scodeggio, M., et al. 1998, A\&A, {\it in
press}(astro-ph/9807336)
\bibitem{stanforda} Stanford, A., Elston, R., Eisenhardt, P., Spinrad, H.,
Stern, D. \&  Dey, A., 1997, AJ, 114, 2232
\bibitem{stanfordb} Stanford, A., Eisenhardt, P. \& Dickinson, M., 1998, ApJ,
492, 461
\end{thebibliography}
\end{document}